\documentclass[doublecol,linenumbers]{epl2}
\usepackage{amsmath}
\usepackage{graphicx}
\usepackage{dsfont}

\begin{document}

\title{Topological phase transition from nodal to nodeless d-wave
superconductivity in electron-doped cuprate superconductors}
\author{Guo-Yi Zhu\inst{1} \and Guang-Ming Zhang\inst{1,2}}
\date{\today}

\institute{
\inst{1}State Key Laboratory of Low-Dimensional Quantum Physics and Department of Physics, Tsinghua University, Beijing 100084, China. \\
\inst{2}Collaborative Innovation Center of Quantum Matter, Beijing 100084, China.}

\abstract
{Unlike the hole-doped cuprates, both nodal and nodeless superconductivity
(SC) are observed in the electron-doped cuprates. To understand these two
types of SC states, we propose a unified theory by considering the
two-dimensional t-J model in proximity to an antiferromagnetic (AF)
long-range ordering state. Within the slave-boson mean-field
approximation, the d-wave pairing symmetry is still the most energetically
favorable even in the presence of the external AF field. In the nodal phase,
it is found that the nodes carry vorticity and are protected by the adjoint
symmetry of time-reversal and one unit lattice translation. Robust edge
modes are obtained, suggesting the nodal d-wave SC being a topological
weak-pairing phase. As decreasing the doping concentration or increasing the
AF field, the nodes with opposite vorticity annihilate and the nodeless
strong-pairing phase emerges. The topological phase transition is
characterized by a critical point with anisotropic Bogoliubov
quasiparticles, and a universal understanding is thus established for all
electron-doped cuprates.}
\maketitle

\section{Introduction}

Over thirty-year-effort, the consensus has been reached that the rich phase
diagram in cuprate superconductors mainly arises from the strong electronic
correlations\cite{AndersonLeeRiceZhang,LeeNagaosaWen}. The low-energy theory
of the doped Mott insulator is believed to be captured by the single band
t-J model\cite{ZhangRice}. There have been tremendous studies on the t-J
model\cite{KotliarLiu,Gros88}, most of which show d-wave pairing
superconductivity (SC), which has been confirmed by hole-doped cuprates\cite%
{Shen,Wollman,Tsuei}. However, a remarkable asymmetry exists between hole
doping (p-type) and electron doping (n-type) cuprates\cite{Armitage}. One of
the most studied n-type family Re$_{2-x}$Ce$_{x}$CuO$_{4}$ (Re is a
trivalent rare-earth cation) displays antiferromagnetic (AF) long-range
order up to a relatively high dopant concentration $0.14$ (Ref.\cite%
{Yamada,Motoyama}) before the nodal d-wave SC appears. In contrast, the
other n-type family A$_{1-x}$La$_{x}$CuO$_{2}$ (A=Sr,Ca) compound has a
nodeless SC gap, and the d-wave pairing symmetry is suspected\cite%
{Chen,Liu,White,Khasanov}.

Recently, by virtue of angular resolved photoemission spectroscopy
measurement on the epitaxially stabilized Sr$_{1-x}$La$_{x}$CuO$_{2}$ thin
films by oxide molecular-beam epitaxy, Harter et al. observed that the Fermi
surface of the SC samples consists of a large electron pocket around $(\pi
,0)$ and a tiny hole pocket surrounding $(\pi /2,\pi /2)$, which perfectly
fits into a tight-binding electronic band structure with an AF long-range
order\cite{Harter}. Moreover, in the SC state, a strong coupling between the
charge carriers and the AF long-range order can push the nodal
quasiparticles below the Fermi level, leading to nodeless d-wave SC without
a change in the pairing symmetry of the order parameter\cite{Harter}.
Actually, such a feature had been also noticed in the single crystal Re$%
_{2-x}$Ce$_{x}$CuO$_{4}$ samples\cite{Armitage2001,Matsui2005,Park2007}.

In the present paper, we will carefully study the Fermi surface evolution as
varying the doping concentration and examine the SC pairing symmetry in the
two-dimensional t-J model in the presence of a staggered magnetic field,
which mimics the AF long-range correlations in the n-type family cuprates%
\cite{Yuan2006,Bansil2007}. In real materials, the dopant is very likely to
be inhomogenious, resulting in the superconducting regions in proximity to
some AF regions. With the slave-boson mean-field (MF)
approximation in the hole picture, the Fermi surface is composed of a large
electron pocket around $(\pi ,0)$ and a tiny hole pocket surrounding $(\pi
/2,\pi /2)$, which gradually emerges as increasing the doping concentration.
In the superconducting phase, we found that the d-wave pairing symmetry is
the most energetically favorable even in the presence of strong AF field. In
the nodal phase, the external AF field duplicates the nodes via the AF
scattering process, which does not change the nature of the nodal d-wave SC
phase. Moreover, the nodes carry vorticity $\pm \revision{2}$ and are
protected by adjoint symmetry $\tilde{\mathcal{T}}$ of the time-reversal and
one unit lattice translation, and the corresponding robust edge states are
also obtained. As increasing the AF field or decreasing the doping level,
the nodes with opposite vorticity annihilate, and the nodal low-energy
excitations are gapped out, leading to a nodeless d-wave SC. A topological
phase transition occurs between the nodal phase and the nodeless phase. The
critical point is characterized by anisotropic Bogoliubov quasiparticles
with quadratic dispersions along the nodal direction and the Dirac line
dispersions perpendicular to the nodal direction. We note that distinct from the topological transition of nodal d-wave induced
via additional spin-orbital interaction\cite{TsueiM}, the topological phase transition we
address here is intrinsic to the electron doped cuprate materials.

\section{Model and Theory}

Since the doped electrons of the cuprates reside at the Cu $3d_{x^{2}-y^{2}}$
orbital forming the full $3d^{10}$ configuration, the basic physics is well
captured by the t-J model on a two-dimensional square lattice\cite%
{LeeNagaosaWen}. Experimental measurements suggest that the nearest neighbor
(n.n.), the next nearest neighbor (n.n.n.), and the next next nearest
neighbor (n.n.n.n.) hopping should be taken into account. By including an
external AF field, the model Hamiltonian is defined by
\begin{eqnarray}
H &=&t\sum_{\mathbf{r\delta }\sigma }c_{\mathbf{r,}\sigma }^{\dagger }c_{%
\mathbf{r}+\mathbf{\delta },\sigma }+t^{\prime }\sum_{\mathbf{%
r\gamma }\sigma }c_{\mathbf{r},\sigma }^{\dagger }c_{\mathbf{r}+\mathbf{%
\gamma },\sigma }-\mu _{0}\sum_{\mathbf{r}}n_{\mathbf{r}}  \notag \\
&&+t^{\prime \prime }\sum_{\mathbf{r\delta }\sigma }c_{\mathbf{r}%
,\sigma }^{\dagger }c_{\mathbf{r}+2\mathbf{\delta },\sigma }+m_{s}\sum_{%
\mathbf{r}\sigma }\sigma e^{i\mathbf{Q\cdot r}}c_{\mathbf{r},\sigma
}^{\dagger }c_{\mathbf{r},\sigma }  \notag \\
&&+\frac{J}{2}\sum_{\mathbf{r,\delta }}\left( \mathbf{S}_{\mathbf{r}}\cdot
\mathbf{S}_{\mathbf{r}+\mathbf{\delta }}-\frac{1}{4}n_{\mathbf{r}}n_{\mathbf{%
r}+\mathbf{\delta }}\right) ,
\end{eqnarray}%
where \textbf{$\delta $} and \textbf{$\gamma $} denote the n.n. and n.n.n.
vectors, respectively, and $m_{s}$ is the external AF field with wave vector
$\mathbf{Q=}\left( \pi ,\pi \right) $. The model is subject to a local
constraint $\sum_{\sigma }c_{\mathbf{r},\sigma }^{\dagger }c_{\mathbf{r}%
,\sigma }\preceq 1$. The band parameters are adopted by $t=215$ meV, $%
t^{\prime }=-0.16t$, $t^{\prime \prime }=0.2t$, $J=0.3t$, which
are relevant to experiments\cite{Harter}. Note that this model is expressed
in the hole representation so that $c_{\mathbf{r},\sigma }^{\dagger }$
creates a hole in the Cu $3d_{x^{2}-y^{2}}$ orbital and therefore
the hopping parameters differ from the hole doped cases by a
sign\cite{CSTing}. Accordingly, in what follows the hole/electron pocket as
seen from the band structure should correspond to the electron/hole pocket
as observed by experiments, and the pocket we address in the following will
refer to the experimental pocket to save confusion.

Firstly we would like to point out that there is a special symmetry in the
model Hamiltonian. Since the external AF field breaks the lattice
translation and time-reversal symmetries, the square lattice is divided into
two sublattice A and B ($r_{x}+r_{y}=even$ and $r_{x}+r_{y}=odd$) that are
subject to opposite magnetic fields. The sublattice degrees of freedom span
a two dimensional space in terms of Pauli operator $\tau _{x}$: $\tau _{x}=1$
stands for the sublattice A and $\tau _{x}=-1$ for the sublattice B. It is
obvious that one unit lattice translation operation is equivalent to
applying the operator $\tau _{z}$, which flips the sublattice degree of
freedom. It can be proved that the coupling of the electrons with the AF
field commutes with an adjoint operator $\tilde{\mathcal{T}}\equiv \mathcal{T%
}\tau _{z}$, where $\mathcal{T}\equiv i\sigma _{y}\mathcal{K}$ is the
time-reversal operator and $\tau _{z}$ is the one unit lattice translational
operator. Therefore, the model Hamiltonian has the adjoint $\tilde{\mathcal{T%
}}$ symmetry, which is much similar to the time-reversal symmetry in such an
adjoint way due to the zero net magnetization in each unit cell.

To treat the local constraint, we employ the slave-boson
decomposition: $c_{\mathbf{r},\sigma }=b_{\mathbf{r}}^{\dagger }f_{\mathbf{r}%
,\sigma }$ and the constraint is rewritten into $b_{\mathbf{r}}^{\dagger }b_{%
\mathbf{r}}+\sum_{\sigma }f_{\mathbf{r},\sigma }^{\dagger }f_{\mathbf{r}%
,\sigma }=1$, which can be enforced by introducing a Lagrangian multiplier $%
\lambda $. When the holons condense $\left\langle b_{\mathbf{r}%
}\right\rangle =\left\langle b_{\mathbf{r}}^{\dagger }\right\rangle =\sqrt{x}
$, the fermionic spinon parts are left. To decouple the spin superexchange
term, MF order parameters are introduced
\begin{eqnarray}
\kappa &\equiv &\frac{J}{4}\left\langle f_{\mathbf{r},\uparrow }^{\dagger
}f_{\mathbf{r}+\mathbf{\delta },\uparrow }+f_{\mathbf{r},\downarrow
}^{\dagger }f_{\mathbf{r}+\mathbf{\delta },\downarrow }\right\rangle ,
\notag \\
\Delta _{\mathbf{\delta }} &\equiv &\frac{J}{4}\left\langle f_{\mathbf{r}%
,\uparrow }f_{\mathbf{r}+\mathbf{\delta },\downarrow }-f_{\mathbf{r}%
,\downarrow }f_{\mathbf{r}+\mathbf{\delta },\uparrow }\right\rangle .
\end{eqnarray}%
In generally we assume $\Delta _{x}=\Delta _{d}+i\Delta _{s}$ and $\Delta
_{y}=-\Delta _{d}+i\Delta _{s}$, where $\Delta _{s}$ and $\Delta _{d}$ are
amplitudes of n. n. spin-singlet pairing with $s_{x^{2}+y^{2}}$- and $%
d_{x^{2}-y^{2}}$-symmetries, respectively. Then in momentum space the MF
Hamiltonian can be written as%
\begin{eqnarray}
H_{\text{mf}} &=&\sum_{\mathbf{k\sigma }}\left[ \left( \epsilon _{\mathbf{k}%
}+\epsilon _{\mathbf{k}}^{\prime }-\mu \right) f_{\mathbf{k},\sigma
}^{\dagger }f_{\mathbf{k},\sigma }+m_{s}\sigma f_{\mathbf{k},\sigma
}^{\dagger }f_{\mathbf{k+Q},\sigma }\right]  \notag \\
&&+\sum_{\mathbf{k}}\left( \Delta _{\mathbf{k}}f_{\mathbf{k},\uparrow
}^{\dagger }f_{-\mathbf{k},\downarrow }^{\dagger }+\Delta _{\mathbf{k}%
}^{\ast }f_{-\mathbf{k},\downarrow }f_{\mathbf{k},\uparrow }\right) ,
\label{H_mf}
\end{eqnarray}%
where
\begin{eqnarray*}
\epsilon _{\mathbf{k}} &\equiv &2(tx+\kappa )\left( \text{cos}k_{x}+\text{cos%
}k_{y}\right) ,\mu \equiv \mu _{0}-\lambda , \\
\epsilon _{\mathbf{k}}^{\prime } &\equiv &4t^{\prime }x\text{cos}k_{x}\text{%
cos}k_{y}+2t^{\prime \prime }x\left( \text{cos}2k_{x}+\text{cos}%
2k_{y}\right) , \\
\Delta _{\mathbf{k}} &\equiv &2\Delta _{d}\left( \text{cos}k_{x}-\text{cos}%
k_{y}\right) +i2\Delta _{s}\left( \text{cos}k_{x}+\text{cos}k_{y}\right) .
\end{eqnarray*}%
Note that the kinetic energy part is renormalized by the doping
concentration but the external AF field $m_{s}$ is not, because
the AF field in fact couples only to the spinon. This suggests that a small
AF field $m_{s}$ can have a significant effect on the band structure and SC
pairing. One important point is that the MF Hamiltonian still preserves the
adjoint symmetry $\tilde{\mathcal{T}}$.

The MF Hamiltonian can be diagonalized in two steps. First, the normal state
band structure determined by the first two terms of Eq.(\ref{H_mf}) can be
derived as
\begin{equation}
\xi _{\pm ,\mathbf{k}}=\epsilon _{\mathbf{k}}^{\prime }-\mu \pm \sqrt{%
\epsilon _{\mathbf{k}}^{2}+m_{s}^{2}},  \label{quasiSpc}
\end{equation}%
in the quasiparticles of%
\begin{eqnarray}
\psi _{+,\mathbf{k},\sigma }^{\dagger } &=&\left( \text{cos}\theta _{\mathbf{%
k}}\right) f_{\mathbf{k},\sigma }^{\dagger }+\sigma \left( \text{sin}\theta
_{\mathbf{k}}\right) f_{\mathbf{Q}+\mathbf{k},\sigma }^{\dagger },  \notag \\
\psi _{-,\mathbf{k},\sigma }^{\dagger } &=&\left( \text{sin}\theta _{\mathbf{%
k}}\right) f_{\mathbf{k},\sigma }^{\dagger }-\sigma \left( \text{cos}\theta
_{\mathbf{k}}\right) f_{\mathbf{Q}+\mathbf{k},\sigma }^{\dagger },
\label{slave2quasi}
\end{eqnarray}%
where $\theta _{\mathbf{k}}\equiv \frac{1}{2}\tan ^{-1}\frac{m_{s}}{\epsilon
_{\mathbf{k}}}\in \left[ 0,\frac{\pi }{2}\right] $. The normal state
spectrum has two-fold Kramer's degeneracy due to the $\tilde{\mathcal{T}}$
antiunitary symmetry, then the adjoint operation of $\mathcal{\tilde{T}}$ on
the original fermions is precisely equivalent to the time-reversal $\mathcal{%
T}$ acting on the quasiparticles, ie.,
\begin{equation}
\mathcal{\tilde{T}}^{-1}\left(
\begin{array}{c}
f_{\mathbf{k},\sigma } \\
f_{\mathbf{k}+\mathbf{Q},\sigma }%
\end{array}%
\right) \tilde{\mathcal{T}}\text{ }\Longleftrightarrow \mathcal{T}%
^{-1}\left(
\begin{array}{c}
\psi _{+,\mathbf{k},\sigma } \\
\psi _{-,\mathbf{k},\sigma }%
\end{array}%
\right) \mathcal{T}\text{ .}
\end{equation}%
Physically, this is because the sublattice degrees of freedom are embedded
within the quasiparticles.

The next step is to turn on the superconducting pairing between normal
quasiparticles. When the pairing matrix is written in the quasiparticle
Nambu spinor,
\begin{equation*}
\Psi _{\pm ,\mathbf{k}}^{\dagger }\equiv \left(
\begin{array}{cccc}
\psi _{\pm ,\mathbf{k},\uparrow }^{\dagger } & \psi _{\pm ,\mathbf{k}%
,\downarrow }^{\dagger } & \psi _{\pm ,-\mathbf{k},\downarrow } & -\psi
_{\pm ,-\mathbf{k},\uparrow }%
\end{array}%
\right) ,
\end{equation*}%
the interband pairing is found to be absent, owing to the n. n. pairing and
singlet pairing nature. As a result, the two species of quasiparticles are
decoupled even in pairing channel i.e. $H_{\text{mf}}=\frac{1}{4}\sum_{%
\mathbf{k},\alpha =\pm }\Psi _{\alpha ,\mathbf{k}}^{\dagger }H_{\alpha }(%
\mathbf{k})\Psi _{\alpha ,\mathbf{k}}$, where
\begin{equation}
H_{\pm }(\mathbf{k})=\xi _{\pm ,\mathbf{k}}\rho _{z}\sigma_0\pm
(\Delta _{\mathbf{k}}\rho _{+}+\Delta _{\mathbf{k}}^{\ast }\rho _{-})%
\sigma_0,
\end{equation}%
where $\rho _{+}$, $\rho _{-}$, and $\rho _{z}$ denote three 2$\times $2
Pauli matrices acting in the particle-hole sector. It is straightforward to
obtain the Bogoliubov quasiparticle spectrum%
\begin{equation}
E_{\pm ,\mathbf{k}}=\sqrt{\xi _{\pm ,\mathbf{k}}^{2}+|\Delta _{\mathbf{k}%
}|^{2}}.  \label{BogoSpc}
\end{equation}%
The Bogoliubov quasi-particles are given by
\begin{eqnarray*}
\eta _{\pm ,\mathbf{k},\sigma }^{\dagger } &=&\left( \cos \beta _{\pm ,%
\mathbf{k}}\right) \psi _{\pm ,\mathbf{k},\sigma }^{\dagger }\pm \sigma
\left( \sin \beta _{\mathbf{k,}\pm }\right) e^{\text{i$\phi $}_{\mathbf{k}%
}}\psi _{\pm ,-\mathbf{k},-\sigma }, \\
\eta _{\pm ,-\mathbf{k},\sigma } &=&\left( \sin \beta _{\pm ,\mathbf{k}%
}\right) \psi _{\pm ,\mathbf{k},\sigma }^{\dagger }\mp \sigma \left( \cos
\beta _{\mathbf{k,}\pm }\right) e^{\text{i$\phi $}_{\mathbf{k}}}\psi _{\pm ,-%
\mathbf{k},-\sigma },
\end{eqnarray*}%
where $\beta _{\mathbf{k,}\pm }\equiv \frac{1}{2}\tan ^{-1}\frac{\left|
\Delta _{\mathbf{k}}\right| }{\xi _{\pm ,\mathbf{k}}}\in \left[ 0,\frac{\pi
}{2}\right] $ and $\phi _{k}\equiv -\tan ^{-1}\frac{\text{Im}\Delta _{%
\mathbf{k}}}{\text{Re$\Delta $}_{\mathbf{k}}}$.

By filling out the negative energy states, the ground-state energy density
can be written as
\begin{equation*}
\varepsilon _{g}=-\int \frac{dk_{x}dk_{y}}{8\pi ^{2}}\left( E_{+,\mathbf{k}%
}+E_{-,\mathbf{k}}\right) -\mu x+\frac{8}{J}\left( \kappa ^{2}+\Delta
_{s}^{2}+\Delta _{d}^{2}\right).
\end{equation*}
Then the saddle point equations can be derived by minimizing the ground
state energy $\partial \varepsilon _{g}/\partial (\kappa ,\Delta _{s},\Delta
_{d},\mu )=0$, from which the MF parameters $(\kappa ,\Delta _{s},\Delta
_{d},\mu )$ are determined self-consistently. In terms of Bogoliubov
quasiparticles, the imaginary-time Green's function $G_{\sigma }(\mathbf{k}%
,\tau )=-\left\langle T_{\tau }f_{\mathbf{k},\sigma }(\tau )f_{\mathbf{k}%
,\sigma }^{\dagger }(0)\right\rangle $ can be deduced to
\begin{eqnarray}
&&G_{\sigma }\left( \mathbf{k},i\omega _{n}\right) =\left( \cos ^{2}\theta _{%
\mathbf{k}}\right) \left( \frac{\cos ^{2}\beta _{\mathbf{k,}+}}{i\omega
_{n}-E_{+,\mathbf{k}}}+\frac{\sin ^{2}\beta _{\mathbf{k,}+}}{i\omega
_{n}+E_{+,\mathbf{k}}}\right)  \notag \\
&&\text{ \ \ \ \ \ \ \ \ \ }+\left( \sin ^{2}\theta _{\mathbf{k}}\right)
\left( \frac{\cos ^{2}\beta _{\mathbf{k,}-}}{i\omega _{n}-E_{-,\mathbf{k}}}+%
\frac{\sin ^{2}\beta _{\mathbf{k,}-}}{i\omega _{n}+E_{-,\mathbf{k}}}\right) ,
\end{eqnarray}%
and the corresponding spectral function $A\left( \mathbf{k},\omega \right) =-%
\frac{1}{\pi }$Im$G_{\sigma }\left( \mathbf{k},\omega +i0^{+}\right) $ is
thus obtained,%
\begin{eqnarray}
&&A\left( \mathbf{k},\omega \right) =\frac{1}{4}\sum_{\alpha =\pm }\left(
1+\alpha \frac{\epsilon _{\mathbf{k}}}{\sqrt{\epsilon _{\mathbf{k}%
}^{2}+m_{s}^{2}}}\right)  \notag \\
&&\text{ \ \ \ \ \ \ \ \ \ \ \ \ \ }\times \left[ \left( 1+\frac{\xi
_{\alpha ,\mathbf{k}}}{E_{\alpha ,\mathbf{k}}}\right) \delta \left( \omega
-E_{\alpha ,\mathbf{k}}\right) \right.  \notag \\
&&\text{ \ \ \ \ \ \ \ \ \ \ \ \ \ \ }\left. +\left( 1-\frac{\xi _{\alpha ,%
\mathbf{k}}}{E_{\alpha ,\mathbf{k}}}\right) \delta \left( \omega +E_{\alpha ,%
\mathbf{k}}\right) \right] .
\end{eqnarray}%
In the following we perform the numerical calculations to solve the saddle
point equations and analyze the Fermi surface structure in the normal state
and the Bogoliubov quasiparticle spectrum.

\section{Fermi surface and phase diagram}

\begin{figure}[t]
\par
\includegraphics[width=7.8cm]{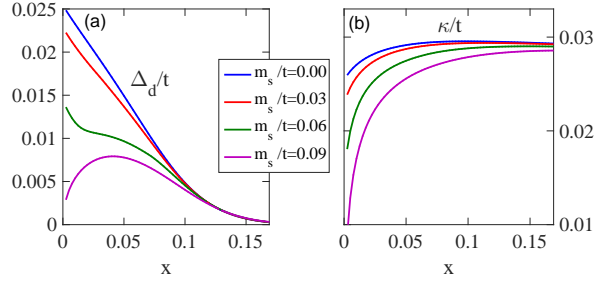}
\caption{For given values of the external AF field, the d-wave pairing
amplitude (a) and the valence bond parameter (b) as the function of the
doping concentration.}
\label{Fig1}
\end{figure}

In the reasonable parameter regime the numerical calculations always found $%
\Delta _{s}=0$ and the d-wave pairing is energetically favorable, i.e., $%
\Delta _{d}\neq 0$. For given the values of $m_{s}/t=0$, $0.03$, $0.06$, and
$0.09$, the parameters $\Delta _{d}$ and $\kappa $ as a function of the
doping concentration are shown in Fig.\ref{Fig1}. The d-wave pairing
amplitude roughly decreases with doping and tends to vanish beyond $17$\%
doping level, however, the AF field $m_{s}$ suppresses the pairing
amplitude. The valence bond parameter $\kappa $ has the same sign with n.n.
hopping and its value increases with doping and decreases with the field $%
m_{s}$.

\begin{figure}[t]
\par
\includegraphics[width=7.8cm]{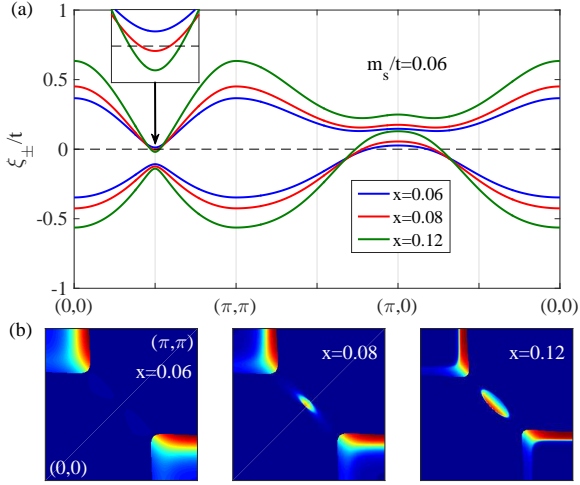}
\caption{(a) The band structure in the external AF field $m_{s}/t=0.06$ for
given doping concentration $x=0.06$, $0.08$, $0.12$. The inset shows the
enlarged feature around the nodal point $\mathbf{K}$. (b) The corresponding
intensity map around the Fermi surface obtained from the electron spectral
function.}
\label{Fig2}
\end{figure}

Next let us turn off the pairing amplitude to analyze the Fermi surface
structure and its evolution for given doping concentrations $x=0.06$, $0.08$%
, $0.12$ and a reasonable AF field $m_{s}/t=0.06$. As we shown previously,
two quasiparticle bands $\xi _{\pm ,\mathbf{k}}$ are resulted from the band
folding and split by the AF field. Away from half filling, the doped
electrons enter into the electron pocket around ($\pi $,$0$), which is
gradually enlarged by increasing the doping level. On the other hand, since
the effective n.n. hopping depends linearly on the doping ratio, increasing
the doping level also enlarges the band width and makes the upper band $\xi
_{+,\mathbf{k}}$ approach towards until crossing the Fermi level, resulting
in a tiny hole pocket around $\mathbf{K=}\left( \pi /2,\pi /2\right) $. In
Fig.\ref{Fig2}a, as increasing the doping concentration $x$, the upper band $%
\xi _{+,\mathbf{k}}$ comes downwards to touch the Fermi level around the $%
\mathbf{K}$ point, while the lower one $\xi _{-,\mathbf{k}}$ crosses the
Fermi level around the $(\pi ,0)$ point. The corresponding Fermi surfaces
can be shown by the intensity map by integrating out the spectral function
near the Fermi level $\int_{-0.06}^{0}A(\mathbf{k},\omega )\,d\omega $,
displayed in Fig.\ref{Fig2}b. Both the hole and electron pockets grow with
increasing doping level and tend to be connected, reconstructing a large
Fermi-surface. Such a feature qualitatively agrees with the experimental
observation\cite{Harter}. Note that with a different value of AF field the
feature does not change drastically; for example, stronger AF field would
only require larger dopant concentration for the hole pocket around $\mathbf{%
K}$ to occur, and the evolution is similar.

\begin{figure}[t]
\par
\includegraphics[width=7.8cm]{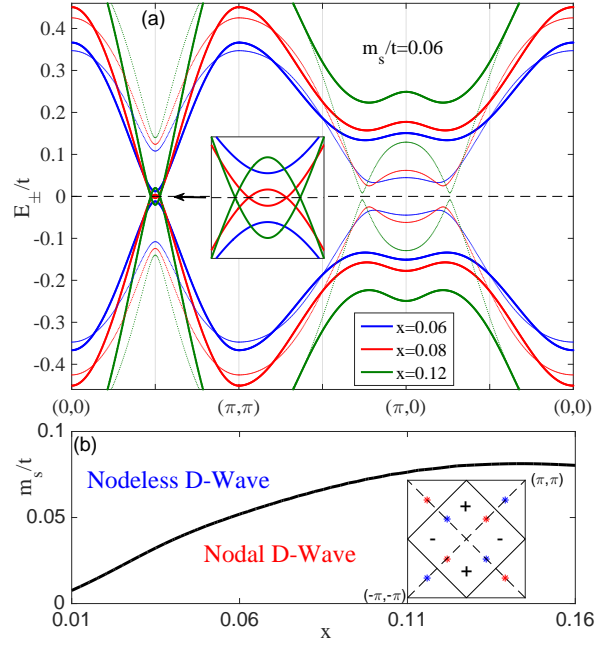}
\caption{(a) The spectrum of Bogoliubov quasiparticles evolves with
increasing the doping level for a fixed AF field. The solid line denotes $%
\pm E_{+,\mathbf{k}}$ while the dashed line for $\pm E_{-\mathbf{,k}}$. The
nodes appear around the point $\mathbf{K}=(\protect\pi /2,\protect\pi /2)$,
and the insert shows the enlarged feature. (b) The superconducting phase
diagram. The insert shows the positions of nodes in the Brillouin zone,
where $\pm$ represents the sign of pairing field, and dots in red and blue
colors represent nodes carrying $\pm 2$ vorticity, respectively.}
\label{Fig3}
\end{figure}

In the SC phase, the Bogoliubov quasiparticle spectra $E_{\pm ,\mathbf{k}}$
are displayed in Fig.\ref{Fig3}a for given dopant concentrations $x=0.06$, $%
0.08$, and $0.12$ and the reasonable AF field $m_{s}/t=0.06$. Due to the
presence of the d-wave pairing, an energy gap opens up in the lower
Bogoliubov quasiparticle band $E_{-,\mathbf{k}}$ around the point $(\pi ,0)$%
, making the Fermi pocket around the anti-nodal point fully gapped. The
low-energy excitations are from the upper Bogoliubov quasiparticle band $%
E_{+,\mathbf{k}}$ around the nodal point. Moreover, we also found that the
nodal Fermi pocket preserves a pair of nodes residing at the symmetric
points $\mathbf{K}_{\pm }$ above a critical doping concentration, below
which the nodes are also gapped out. The full phase diagram is calculated
and shown in Fig.\ref{Fig3}b.

\section{Topological properties of the nodal d-wave SC}

In the nodal d-wave SC phase, the nodes exist and only exist at the
intersections of the Fermi pocket and the zero line of d-wave pairing field,
i.e., $\xi _{+}(k_{x},k_{y})=0$. There are eight nodes in the first
Brillouin zone, as schematically shown in the insert of Fig.\ref{Fig3}b, but
only two of them are independent, denoted as $\mathbf{K}_{\pm }\equiv \left(
K_{\pm },K_{\pm }\right) $ within the first quadrant Brillouin zone ($%
0<K_{+}<\pi /2$ and $K_{-}=\pi -K_{+}>\pi /2$) . The other six nodes are
merely the reflection images of $\mathbf{K}_{\pm }$ and can be neglected for
the moment. So when the gapped quasiparticle states are neglected, the
low-energy effective Hamiltonian contains two nodal valleys that can be
obtained by expanding $H_{+}(\mathbf{k})=(\xi _{+,\mathbf{k}}\rho
_{z}+\Delta _{\mathbf{k}}\rho _{x})\sigma_0$ around the nodal
points $\mathbf{K}_{\pm }$. We thus have
\begin{equation}
H_{\text{eff}}\left( \mathbf{K}_{\pm }+\mathbf{q}\right) =(\pm
v_{3}q_{+}\rho _{z}+v_{1}q_{-}\rho _{x})\sigma_0\equiv (\mathbf{h}%
_{\pm }(\mathbf{q})\cdot \mathbf{\rho)\sigma_0 },
\label{NodalEffHam}
\end{equation}%
which is written as two copies of two-dimensional Dirac
Hamiltonian with canonical coordinates $q_{\pm }\equiv q_{x}\pm q_{y}$ and
two characteristic velocities: $v_{1}=-2\Delta _{d}\sin K_{+}$ and
\begin{eqnarray*}
&&v_{3}=-2\left( t^{\prime }x+2t^{\prime \prime }x\right) \sin
2K_{+} \\
&&\text{ \ \ \ \ \ \ }-\frac{4(tx+\kappa )^{2}\sin 2K_{+}}{\sqrt{%
m_{s}^{2}+16(tx+\kappa )^{2}\cos ^{2}K_{+}}}.
\end{eqnarray*}%
This two-dimensional Dirac type Hamiltonian resembles a ''magnetic field'' $%
\mathbf{h}_{\pm }(\mathbf{q})$ coupling the Nambu spinor in the momentum
space ($h_{\pm }^{x}=v_{1}q_{-}$,$h_{\pm }^{y}=0$, $h_{\pm }^{z}=$ $\pm
v_{3}q_{+}$). The magnitude of the ''magnetic field'' $\left| \mathbf{h}%
_{\pm }(\mathbf{q})\right| $ determines the energy spectrum for the
low-energy excitations, while its unit direction $\mathbf{n}_{\pm }(\mathbf{q%
})\equiv \mathbf{h}_{\pm }(\mathbf{q})/\left| \mathbf{h}_{\pm }(\mathbf{q}%
)\right| $ is responsible for pinning the ground state. Moreover, the nodes $%
\mathbf{K}_{\pm }$ turn out to be the core of vortices, whose vorticity can
be calculated by the topological winding number
\begin{equation}
w_{\pm }=2\oint \frac{d\mathbf{q}}{2\pi }\cdot \left[ n_{\pm
}^{x}(\mathbf{q})\nabla _{\mathbf{q}}n_{\pm }^{z}(\mathbf{q})-n_{\pm }^{z}(%
\mathbf{q})\nabla _{\mathbf{q}}n_{\pm }^{x}(\mathbf{q})\right] ,
\label{windingnum}
\end{equation}%
where the winding number is multiplied by 2 due to spin
degeneracy. It turns out $w_{\pm }=\pm 2$. The vortices cause
the ground-state pairing wave function to experience a nonzero Berry phase
for any closed path surrounding each node in the momentum space. And the
lower dimensional model confined to this loop would be fully gapped and
topologically nontrivial. Consequently, the vorticity implies a nontrivial
topology of the nodal d-wave SC, which is supposed to manifest in the edges
due to bulk edge correspondence.

\begin{figure*}[t]
\par
\includegraphics[width=16cm]{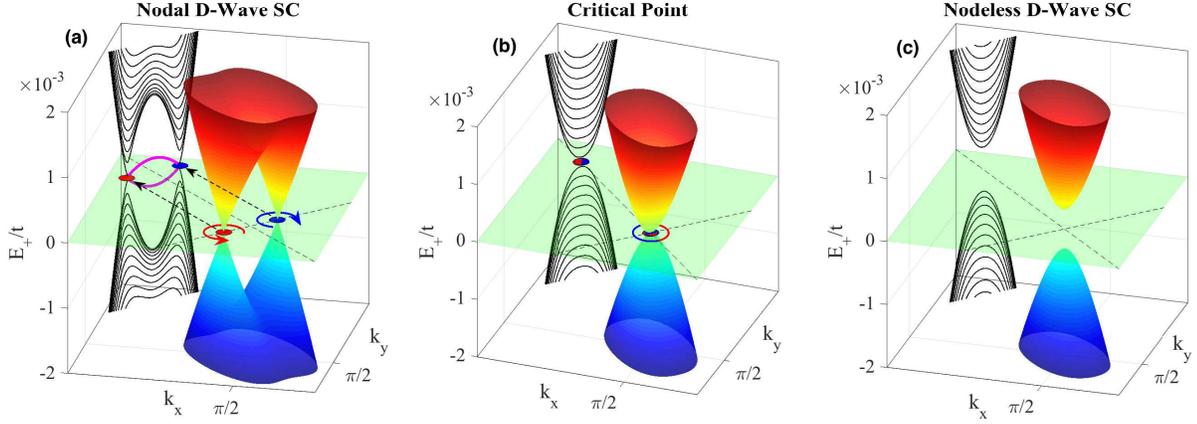}
\caption{For $m_{s}/t=0.06$ and $x=0.075,0.072975,0.0725$ respectively, the
bulk spectrum of closed system are shown in colored sheets, while the
spectrum of lattice in cylinder geometry (size 600$\times $512) with $\left(
1,\bar{1}\right) $ open edges are shown by the black curves behind. (a) the
nodal d-wave SC phase, where a pair of nodes reside symmetrically with
respect to the point $\mathbf{K}=(\protect\pi /2,\protect\pi /2)$; (b) at
the critical point; (c) the nodeless d-wave SC. With decreasing dopant
concentration or increasing the external AF field, the vortex and
anti-vortex approach towards and finally annihilate each other, leading to
the topological phase transition from the nodal d-wave SC to the nodeless
d-wave SC.}
\label{Fig4}
\end{figure*}

For each fixed momentum $k_{(11)}$ along (1,1), the system is effectively a
one dimensional chain along ($1,\bar{1}$). The 1D chain that avoids the
nodes is always fully gapped and characterized by the topological winding
number given by Eq.~(\ref{windingnum}). Any two chains whose $k_{(11)}$
embrace the projection of $\mathbf{K}_{\pm }$ must have their topological
winding number differ by $w_{\pm }=\pm 1$, therefore at least one of them is
topologically nontrivial and would show robust edge modes when ($1,\bar{1}$)
boundary turns open. Then we performed exact diagonalization to the system
in cylinder geometry with ($1,\bar{1}$) open edges, and compare it with the
bulk spectrum given by system of closed boundary. As shown in Fig.\ref{Fig4}%
, the spectrum with ($1,\bar{1}$) open edges is nothing but the spectrum of
the closed system being projected onto the ($1,\bar{1}$) edges, except that
in the nodal d-wave SC phase there appear two additional in-gap edge modes
with $k_{(11)}$ residing on the interval between projection of $\mathbf{K}%
_{\pm }$(corresponding dispersion shown by magenta color curve in Fig.\ref%
{Fig4}a). The in-gap edge modes are topologically protected, which is
analogous to the necessary appearance of Fermi arc on the surface of
three-dimensional Weyl semi-metal. Notice that the ($1,0$) or ($0,1$)
surface edges do not show edge bound states because the vortices and
anti-vortices would collapse upon projection onto those edges. Finally, we
mention that it is the AF field that accounts for the energy splitting of
the topologically protected edge states away from exact zero energy of the $%
d_{x^{2}-y^{2}}$ wave SC in the absence of AF studied by Wang and Lee\cite%
{FaDunghai}.

It is necessary to investigate the robustness of the bulk gap nodes being
subject to all possible perturbations. The low-energy effective Hamiltonian (%
\ref{NodalEffHam}) is a two-dimensional massless Dirac Hamiltonian though it
describes the Bogoliubov excitations rather than the usual U(1) fermionic
excitations. To gap out the nodes is equivalent to endowing extra-mass terms
on the massless model. Regardless of the symmetry requirements, all the
available mass terms are enumerated as the following pairing terms:%
\begin{eqnarray}
&&m_{\pm }\Psi _{\pm ,\mathbf{k}}^{\dagger }\sigma _{0}i\rho _{y}\Psi _{\pm
,-\mathbf{k}};  \notag \\
&&m_{1}\Psi _{\pm ,\mathbf{k}}^{\dagger }\sigma _{x}i\rho _{y}\Psi _{\pm ,-%
\mathbf{k}},\text{ }m_{2}\Psi _{\pm ,\mathbf{k}}^{\dagger }\sigma _{y}\rho
_{y}\Psi _{\pm ,-\mathbf{k}},  \notag \\
&&m_{3}\Psi _{\pm ,\mathbf{k}}^{\dagger }\sigma _{z}i\rho _{y}\Psi _{\pm ,-%
\mathbf{k}}.
\end{eqnarray}%
The first mass term represents the extra singlet pairing that differs from
the existing d-wave pairing field by $\pi /2$ phase, transforming the pure
d-wave pairing symmetry into mixed pairing symmetry, e.g. ($%
d_{x^{2}-y^{2}}+is$), ($d_{x^{2}-y^{2}}+is_{x^{2}+y^{2}}$), ($%
d_{x^{2}-y^{2}}+id_{xy}$). However, the mixed pairing phase inevitably
breaks the adjoint $\tilde{\mathcal{T}}$ symmetry. The other three mass
terms stand for triplet pairing channels without the mirror symmetries,
which are not energetically favorable in the AF spin superexchange coupling
of the t-J model. So the only realistic possible mass term that can directly
gap out the nodes comes from the first extra-mass term, which is
nevertheless forbidden by the adjoint $\tilde{\mathcal{T}}$ symmetry. In
this sense the nodal d-wave SC phase is protected by $\tilde{\mathcal{T}}$.

Moreover, the topology of the nodal d-wave SC also manifests in its
weak-pairing nature. The low-energy sector has a small Fermi pocket around
the nodal points $\mathbf{K}_{\pm }$, and its pairing ground state can be
expressed as
\begin{equation}
|\Omega \rangle \propto \exp \left( \sum_{\mathbf{k}}g_{\mathbf{k}}\psi _{+,%
\mathbf{k},\uparrow }^{\dagger }\psi _{+,-\mathbf{k},\downarrow }^{\dagger
}\right) |\text{FS}\rangle ,
\end{equation}%
which displays a string of poles along the nodal direction inside the
pocket,
\begin{equation}
g_{\mathbf{k}}=-\frac{1-n^{z}(\mathbf{k})}{n^{x}(\mathbf{k})}\propto \frac{1%
}{q_{-}}.
\end{equation}%
This result implies a long tail of the pairing function in real space,
indicating the weak pairing nature of the nodal d-wave SC. Therefore, the
nodal d-wave SC is a weak-pairing topological superconducting phase
protected by the adjoint symmetry $\tilde{\mathcal{T}}$. In contrast, in the
nodeless d-wave SC the Fermi pockets around $\mathbf{K}_{\pm }$ and its
equivalent points are absent, and the pairing function in the momentum space
is analytical. These properties suggest that the nodeless d-wave SC is a
strong-pairing phase corresponding to BEC limit and thus topologically
trivial.

\section{Topological phase transition}

In the presence of the symmetry $\mathcal{\tilde{T}}$, the nodes in the
nodal d-wave phase are protected. The only way to kill the nodes is to let
the vortex-antivortex pairs annihilate each other, which can be done by
gradually increasing the AF field or deceasing the dopant concentration. The
phase transition from weak-pairing nodal d-wave SC to strong-pairing
nodeless d-wave SC is characterized by two nodes with opposite vorticity
merging together at the point $(\pi /2,\pi /2)$.

\begin{figure}[t]
\par
\includegraphics[width=7.8cm]{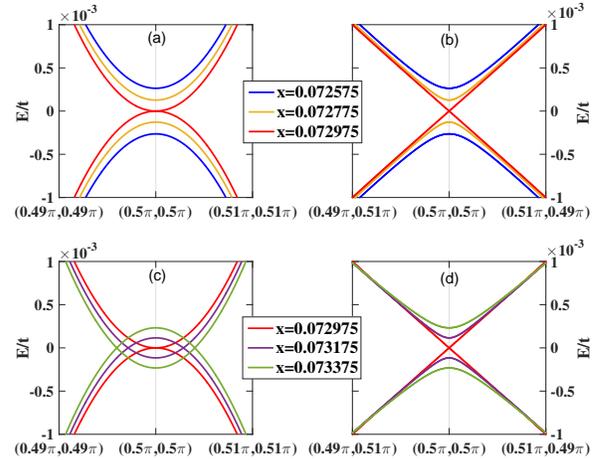}
\caption{Evolution of low-energy spectrum of the Bogoliubov quasi-particles
across phase transition point for a fixed AF field $m_{s}/t=0.06$. (a) and
(c) show the spectra in the nodal direction $k_{x}=k_{y}$. (b) and (d)
display the spectra in the direction $k_{x}=\protect\pi -k_{y}$.}
\label{Fig5}
\end{figure}

To reveal the detailed features of the phase transition, we have to consider
the low-energy excitations near the critical point. Since the effective
Hamiltonian is composed of even functions in the Brillouin zone, we can only
consider one quadrant Brillouin zone and the other areas of the Brillouin
zone are connected by reflections. By expanding $H_{+}(\mathbf{k})=(\xi _{+}(%
\mathbf{k})\rho _{z}+\Delta _{\mathbf{k}}\rho _{x})\sigma_0$
around $\mathbf{K}=(\pi /2,\pi /2)$, the low-energy effective Hamiltonian
can be derived as
\begin{equation}
H_{c}(\mathbf{K}+\mathbf{q})=\left( Aq_{+}^{2}-\mu ^{\prime }\right) \rho
_{z}\sigma_0+vq_{-}\rho _{x}\sigma_0\equiv (\mathbf{h}%
_{c}(\mathbf{q})\cdot \mathbf{\rho})\sigma_0,
\end{equation}%
where
\begin{eqnarray*}
\mu ^{\prime } &=&\mu -m_{s}+4t^{\prime \prime }x,\text{ }%
v=-2\Delta _{d}, \\
A &=&\frac{2(tx+\kappa )^{2}}{m_{s}}+(t^{\prime }+2t^{\prime
\prime })x; \\
h_{c}^{x} &=&vq_{-},\text{ }h_{c}^{y}=0,\text{ }h_{c}^{z}=Aq_{+}^{2}-\mu
^{\prime }.
\end{eqnarray*}%
This quasiparticle spectrum has a linear dispersion along the direction $%
k_{x}=\pi -k_{y}$ but quadratic dispersion along the nodal direction $%
k_{x}=k_{y}$, shown in Fig.\ref{Fig5}. It is clear that the effective
chemical potential $\mu ^{\prime }>0$ stands for the nodal d-wave phase,
while $\mu ^{\prime }<0$ is for the nodeless d-wave phase, so the critical
point is characterized by the effective chemical potential $\mu ^{\prime }=0$%
. In the direction $k_{x}=\pi -k_{y}$, the phase transition can be viewed as
topological phase transition from negative mass to positive mass and
classified by the Z$_{2}$ quantum number. At the critical point, the
low-energy excitations are double-faced, because that the low-energy
excitations of weak pairing phase have massless Dirac spectrum in all
directions while the strong pairing phase shows nonrelativistic spectrum.

\section{Discussion and conclusion}

So far we have discussed the influence of an external AF field on the nodal
d-wave SC. It is important to compare the results with those in the absence
of the AF field, where the nodal d-wave SC has four nodes connected by the
mirror symmetries $M_{x}$ and $M_{y}$ and their mirror partners are shown to
carry opposite vorticity\cite{FaDunghai}. Weak AF field induces the $\mathbf{%
Q}\equiv (\pi ,\pi )$ vector scattering, and creates a copy for all four
nodes. For instance, the node at $\mathbf{K}_{-}-\mathbf{Q}$ induces a node
at $\mathbf{K}_{-}$, whose vorticity differs from the original node because
the pairing field changes sign under the AF vector $\mathbf{Q}$, namely, $%
\Delta _{\mathbf{k}+\mathbf{Q}}=-\Delta _{\mathbf{k}}$. The weak AF field
does not completely destroys the Fermi pocket that enters into the
low-energy sector by crossing the pairing nodal line, and the number of
nodes within magnetic Brillouin zone is still four. Therefore, nodal d-wave
SC phase without the AF field is topologically the same phase as that in the
presence of weak AF field, where nodes are protected by $\mathcal{\tilde{T}}$
symmetry. Only strong enough AF field would drive the vortex-anti-vortex
pairs to annihilate, resulting in the nodeless d-wave SC without breaking $%
\mathcal{\tilde{T}}$ symmetry.

On the other hand, it is necessary to address the possibility to have a
full-gapped topological SC from the symmetry protected nodal d-wave SC.
Similar to the Haldane's approach to realizing a nontrivial Chern insulator
by gapping out the two-dimensional graphene system\cite{HaldaneGraphene}, it
is worth noticing that the gap nodes in the nodal d-wave SC can be directly
gapped out by breaking $\mathcal{\tilde{T}}$ symmetry, which can potentially
lead to nontrivial topological fully gapped SC. There can be two drastically
different ways to introduce mixed singlet pairing channels, depending upon
whether the sign of mass endowed upon the two inequivalent nodes are the
same or the opposite. Introducing the masses $m_{\pm }\Psi _{\pm ,\mathbf{k}%
}^{\dagger }i\rho _{y}\Psi _{\pm ,-\mathbf{k}}$ with the same sign ($%
m_{+}=m_{-}$) would bring the pair of vortex-anti-vortex into merons that
would cancel each other, leading to a trivial full gapped ($%
d_{x^{2}-y^{2}}+is$) SC. On the other hand, introducing masses with the
opposite sign $m_{+}=-m_{-}$ would drive the vortex-anti-vortex pair into
meron and anti-meron which together form a skyrmion, leading to a
topologically full gapped ($d_{x^{2}-y^{2}}+is_{x^{2}+y^{2}}$) SC. Actually,
the extended s-wave pairing plays exactly the role of this nontrivial mass
term in the low-energy Dirac-like Bogoliubov excitations of the nodal d-wave
SC, because its nodal line is along the direction $k_{x}=\pi -k_{y}$. As a
result, with the first quadrant Brillouin zone contributing one unit of
Chern number, we have found that the weak pairing ($%
d_{x^{2}-y^{2}}+is_{x^{2}+y^{2}}$) SC as the valley symmetry protected
topological superconductor can be realized in the hole doped cuprates\cite%
{ZhuZhangWang}.

In summary, we have developed a unified theory to understand both nodal and
nodeless SC observed in the electron-doped cuprates by introducing an
external AF field into the two-dimensional t-J model. Within the slave-%
boson mean-field approximation, the d-wave pairing symmetry is
the most energetically favorable. In the nodal d-wave SC phase, the nodes
are protected by the product of time-reversal and unit lattice translation
symmetries. By increasing the external AF field or decreasing the doping
concentration, the nodes with opposite vorticity annihilate and the nodeless
d-wave SC phase emerges.

\textit{Acknowledgment.- }GMZ would like to thank Ziqiang Wang for his
stimulating discussion and acknowledges the support of National Key Research
and Development Program of China (2016YFA0300300).

\end{document}